\documentclass[review]{elsarticle}
\usepackage{amsmath}
\usepackage{graphics}
\usepackage[utf8]{inputenc}
\usepackage{physics}
\usepackage{hyperref}
\journal{Journal of \LaTeX\ Templates}

\begin{document}

\begin{frontmatter}

\title{Degree of polarization of a spectral electromagnetic Gaussian Schell-model beam passing through 2-f and 4-f lens systems}
\tnotetext[mytitlenote]{Fully documented templates are available in the elsarticle package on \href{http://www.ctan.org/tex-archive/macros/latex/contrib/elsarticle}{CTAN}.}

\author{Rajneesh Joshi, and Bhaskar Kanseri*}
\address{Experimental Quantum Interferometry and Polarization (EQUIP), Department of Physics, Indian Institute of Technology Delhi, Hauz Khas, New Delhi-110016}


\cortext[mycorrespondingauthor]{Corresponding author}
\ead{bkanseri@physics.iitd.ac.in.}


\begin{abstract}
Spectral electromagnetic Gaussian Schell-model (SEGSM) beam is a generalization of Gaussian Schell model beam having parameters with spectral dependence, which offers a basic classical model for random electromagnetic wide-sense statistically stationary beam-like fields. We study degree of polarization (DOP) of a SEGSM beam passing through 2-f and 4-f lens systems. It is observed that for a 2-f lens system, the spectral DOP at the back focal plane of the lens changes with respect to the transverse position from the optic axis, and the spectral parameters of the beam. For a 4-f lens system, the spectral DOP at the back focal plane is independent of the transverse position of the beam, whereas it depends on the beam parameters such as mean value of rms beam-width, rms width of correlation function, and size of aperture placed at the Fourier plane of the lens system.

\end{abstract}
\begin{keyword}
\texttt{}Schell-model beam \sep Vector field\sep Coherence \sep Polarization. 

\end{keyword}
\end{frontmatter}
\section{Introduction}
In classical optics, model sources such as planer, Gaussian Schell-model etc. play an important role in generating light fields having peculiar coherence and statistical features \cite{mandel95, wolf07}. Depending on the nature of the source, the coherence and polarization properties of generated light fields may differ. A basic classical model source for random stationary beam-like fields namely the Gaussian Schell-model (GSM) source is the closest form of the natural quasi-homogeneous light source. In this source, the intensity distribution and the spatial degree of coherence both follow the Gaussian distribution \cite{mandel95}. As a limiting case when the root mean square (rms) width of intensity reduces to zero (infinity), the beams reduce to spherical (plane) waves. These beams were initially introduced for scalar fields, but were later generalized for electromagnetic fields, called as electromagnetic Gaussian Schell-model (EGSM) beams \cite{korotkova04,gori01,gori08}. The coherence and polarization properties of an electromagnetic field is described by a $2\times 2$ electric cross-spectral density (CSD) matrix. For EGSM source, the spectral density of both components of electromagnetic field and the spectral degree of coherence for all field elements (correlations) $xx$, $xy$, $yx$ and $yy$ are having Gaussian distributions. Generally the parameters of EGSM beam, such as rms width of intensity distribution and the degree of coherence are assumed to be constant with wavelength; however in a specific case, they may depend on the wavelength, and then the beam is called as a spectral electromagnetic Gaussian Schell-model (SEGSM) beam \cite{shchepakina13}. GSM sources exhibit rich statistical properties such as coherence and polarization, which makes them applicable in several area of physics such as beam propagation studies, optical communication etc  \cite{zhang17,jian90}. 

Degree of polarization (DOP) of light fields is an important quantity providing a measure of correlations between the orthogonal field components at a single space and time point \cite{mandel95}. By measuring the usual Stokes parameters using a quarter wave plate and a polarizer (called Stokes assembly), one can experimentally determine the DOP for any light field \cite{hauge76,berry77}. The usual Stokes parameters provide information about the state of polarization (SOP) and DOP of light fields \cite{wolf07}. In optics, partially polarized and partially coherent light fields are the most general form of the fields that are having applications in scattering, propagation, communication and in many other areas of science \cite{demos96,cassidy04}. These fields are also known as electromagnetic fields or vector fields. For such fields, the interference manifests both in intensity modulations (intensity fringes) and in polarization modulations \cite{tervo03,kanseri13,setala06a,setala06b}. Recent experimental advances demonstrates that DOP of a light field can be tuned (controlled) from unpolarized (DOP = 0) to fully polarized (DOP = 1) by both amplitude and intensity interferometric techniques \cite{kanseri10,leppanen14,kanseri20}. Recently a new method has been reported to control the DOP of the EGSM sources using 2-f and 4-f lens systems \cite{zhao18}. It was shown that for a EGSM beam passing through a 2-f lens system, DOP at the back focal plane varies at different transverse positions of the beam. Similarly, in a 4-f lens system, if an aperture is placed between the two lenses, the DOP at the output plane depends on the radius of the aperture. However, for an SEGSM beam, since beam properties depend on the wavelength of the source \cite{wang11,shirai05}, it becomes interesting to investigate the polarization features of such GSM beams after passing through the 2-f and 4-f lens configurations.

 In this paper, we aim for a detailed investigation of the  polarization properties of SEGSM beam after passing through 2-f and 4-f lens systems. The obtained theoretical expressions of the SEGSM beam at the back focal plane of 2-f and 4-f lens systems show that the beam parameters such as intensity, beam-width and the correlation function depend on peak wavelength and rms width of these parameters. Through simulations, we have shown the effect of the SEGSM beam parameters on the spectral DOP after passing through both the lens systems. We also observe that the peak value of the spectrum of auto-correlation function and the cross-correlation function affects the spectral DOP of the output beam in both types of the lens systems. 
 
 \section{Spectral electromagnetic Gaussian Schell-model (SEGSM) beam}
 In a Schell-model source, the spectral degree of coherence $\mu(\boldsymbol{\rho}_1,\boldsymbol{\rho}_2,\lambda)$ depends only on the difference of these two points $\boldsymbol{\rho}_1$ and $\boldsymbol{\rho}_2$. 
For a scalar, partially coherent Schell-model source, the cross spectral density (CSD) at two points in space-frequency domain can be written as \cite{mandel95},
 \begin{equation}
 W(\boldsymbol{\rho}_1,\boldsymbol{\rho}_2, \lambda)=\sqrt{S(\boldsymbol{\rho}_1, \lambda)S(\boldsymbol{\rho}_2, \lambda)}\mu(\boldsymbol{\rho}_2-\boldsymbol{\rho}_1,\lambda), \end{equation}
 where $S(\boldsymbol{\rho}_i,\lambda)$ is spectral density at a spatial point ($i=1,2$) for a specific wavelength, and $\mu(\boldsymbol{\rho}_2-\boldsymbol{\rho}_1, \lambda)$ is the spatial degree of coherence which depends on the spatial points through their difference only. In scalar GSM beam, the spectral density and the spatial degree of coherence are Gaussian functions, i.e.,
 \begin{subequations}
  \begin{equation}
 S(\boldsymbol{\rho},\lambda)=S_0(\lambda)exp\left[\frac{-\boldsymbol{\rho}^2}{2\sigma^2}\right],     
  \end{equation}
  \begin{equation}
  \mu(\boldsymbol{\rho}_2-\boldsymbol{\rho}_1,\lambda)=exp\left[\frac{-(\boldsymbol{\rho}_2-\boldsymbol{\rho}_1)^2}{2\delta^2}\right],  
 \end{equation}
 \end{subequations}
where $S_0$ is a positive constant (in general wavelength dependant), $\boldsymbol{\rho}$ denotes the transverse position from the optic axis of the beam, $\sigma$ and $\delta$ are the rms width and rms spatial correlation width of the beam, respectively. In case of a EGSM source, the elements of the $2\times 2$ CSD matrix can be written as
 \begin{equation}
 W_{ij}(\boldsymbol{\rho}_1,\boldsymbol{\rho}_2,\lambda)=\sqrt{S_i(\boldsymbol{\rho}_1,\lambda)S_j(\boldsymbol{\rho}_2,\lambda)}\mu_{ij}(\boldsymbol{\rho}_2-\boldsymbol{\rho}_1,\lambda),    
\end{equation}
 where
 \begin{subequations}
 \begin{equation}
 S_{i}(\boldsymbol{\rho},\lambda)=S_{i0}(\lambda)exp\left[\frac{-\boldsymbol{\rho}^2}{2\sigma_{i}^2}\right],
 \end{equation}
 \begin{equation}
 \mu_{ij}(\boldsymbol{\rho}_{1}, \boldsymbol{\rho}_{2},\lambda)=B_{ij}exp\left[\frac{-(\boldsymbol{\rho}_{2}-\boldsymbol{\rho}_{1})^2}{2\delta_{ij}^2}\right], (i,j=x,y).
 \end{equation}
 \end{subequations}
 In Eq. (4), $S_{i0}$ is maximum spectral density at peak wavelength, $\sigma$ and $\delta$ are the rms beam width and correlation width, respectively for different components ($ij$) of the electric field and $B_{ij}$ denotes the correlation coefficient between the orthogonal field components at a single space and time point. For the SEGSM beam, parameters such as spectral density, rms beam width and rms correlation width depend on wavelength. Since the SEGSM beam is a generalization of the EGSM beam, both the spectral density and spectral degree of coherence of SEGSM beam are also in the Gaussian form \cite{mandel95,wolf07} given as
 \begin{subequations}
 \begin{equation}
 S(\lambda)=S_{0}exp\left[\frac{-(\lambda-\lambda_{S})^2}{2\Lambda_{S}^2}\right],
 \end{equation}
 \begin{equation}
 \sigma(\lambda)=\sigma_{0}exp\left[\frac{-(\lambda-\lambda_{\sigma})^2}{2\Lambda_{\sigma}^2}\right],
 \end{equation}
 \begin{equation}
 \delta(\lambda)=\delta_{0}exp\left[\frac{-(\lambda-\lambda_{\delta})^2}{2\Lambda_{\delta}^2}\right],
 \end{equation}
 \begin{equation}
 \delta_{xy}(\lambda)=\delta_{xy(0)}exp\left[\frac{-(\lambda-\lambda_{\delta_{xy}})^2}{2\Lambda_{\delta_{xy}}^2}\right],
 \end{equation}
\end{subequations}
 where $S_0$, $\sigma_{0}$, $\delta_{0}$, $\delta_{xy(0)}$ are the maximum values of the Gaussian spectrum of spectral density, rms beam width, rms width of auto-correlation function and rms width of cross-correlation function at the peak wavelength $\lambda_n$ and spectral width $\Lambda_n$, [for n= $S$, $\sigma$, $\delta$ and $\delta_{xy}$], respectively. The spectral density and spectral degree of coherence for different electric field components of the SEGSM beam can be expressed as
 \begin{subequations}
 \begin{equation}
 S_{i}(\boldsymbol{\rho},\lambda)=S_i(\lambda)exp\left[\frac{-\boldsymbol{\rho}^2}{2\sigma_{i}^2(\lambda)}\right],
 \end{equation}
 \begin{equation}
 \mu_{ij}(\boldsymbol{\rho}_{2}-\boldsymbol{\rho}_{1},\lambda)=B_{ij}(\lambda)exp\left[\frac{-(\boldsymbol{\rho}_{2}-\boldsymbol{\rho}_{1})^2}{2\delta_{ij}^2(\lambda)}\right],
 \end{equation}
 \end{subequations}
 where $S_i(\lambda)$ denotes the maximum spectral density of the respective spectrum of the component of the beam, $\sigma_{i}(\lambda)$ denotes rms  width of beam, and $\delta_{ij}(\lambda)$ (for $i,j = x,y$) denotes rms width of the correlation function.
 
 \section{Spectral degree of polarization}
 In space frequency domain, the degree of polarization provides correlations between the orthogonal field components at same space-time point for a given wavelength (called spectral DOP). Spectral DOP is defined as \cite{mandel95}
 \begin{equation}
 P(\boldsymbol{\rho},\lambda)=\sqrt{1-\frac{4det W(\boldsymbol{\rho},\boldsymbol{\rho},\lambda)}{[tr W(\boldsymbol{\rho},\boldsymbol{\rho},\lambda)]^2}},    
 \end{equation}
 where $det$ is determinant and $tr$ denotes the trace of the matrix.
The degree of polarization of an EGSM source depends on the rms width of the spectral density of both the components of the electromagnetic field. If these rms widths are same, then the DOP of EGSM beam is independent from that point \cite{mandel95,wolf07}. Owing to the spectral dependence of the beam parameters, the DOP of SEGSM beam needs to be studied carefully.  
\newline
 By substituting beam parameters from Eq. (6) to Eq. (3), we get the elements of CSD matrix for a SEGSM beam as,
 \begin{equation}
 W_{ij}(\boldsymbol{\rho}_{1},\boldsymbol{\rho}_{2},\lambda)=\sqrt{S_{i}(\lambda)S_{j}(\lambda)}B_{ij}(\lambda)exp\left[-(\frac{\boldsymbol{\rho}_{1}^2}{4\sigma_{i}^2(\lambda)}+\frac{\boldsymbol{\rho}_{2}^2}{4\sigma_{j}^2(\lambda)})\right]exp\left[\frac{-(\boldsymbol{\rho}_{2}-\boldsymbol{\rho}_{1})^2}{2\delta_{ij}^2(\lambda)}\right].
\end{equation} 
 The GSM source needs to satisfy a beam realizability condition \cite{mandel95,wolf07,shirai05}.  For the SEGSM beam, the realizability condition is given by \cite{shchepakina13}, 
 \begin{equation}
 \frac{1}{4\sigma_{0}^2}exp\left[\frac{(\lambda-\lambda_{\sigma})^2}{\Lambda_{\sigma}^2}\right]+\frac{1}{\delta_{0}^2}exp\left[\frac{(\lambda-\lambda_{\delta})^2}{\Lambda_{\delta}^2}\right]<<\frac{2\pi^2}{\lambda^2}.
\end{equation} 
Substituting values of Eqs. (5) and (8) into Eq. (7),  we get
 \begin{equation}
 P(\boldsymbol{\rho},\lambda)=\sqrt{1-\frac{4A_1A_2exp-(\frac{\rho^2}{2}[A_3+A_4])}{[A_1exp-(\frac{\rho^2A_3}{2})+A_2exp-(\frac{\rho^2A_4}{2})]^2}[1-|B_{xy}(\lambda)|^2]}.    
 \end{equation}
 Eq. (10) is the general expression for the spectral DOP of a SEGSM beam, where the parameters are given by $A_1=S_{{x}{0}}exp\left[\frac{-(\lambda-\lambda_{{S}_{x}})^2}{2\Lambda^2_{{S}_{x}}}\right]$, $A_2=S_{{y}{0}}exp\left[\frac{-(\lambda-\lambda_{{S}_{y}})^2}{2\Lambda^2_{{S}_{y}}}\right]$, $A_3=\frac{1}{\sigma^2_{{x}{0}}}exp\left[\frac{({\lambda-\lambda_{{\sigma}_x})^2}}{\Lambda^2_{{\sigma}_{x}}}\right]$ and $A_4=\frac{1}{\sigma^2_{{y}{0}}}exp\left[\frac{({\lambda-\lambda_{{\sigma}_y})^2}}{\Lambda^2_{{\sigma}_{y}}}\right]$, respectively. These parameters represent the spectral dependence of spectral density and rms beam width. If the beam parameters are identical for the orthogonal field components, i.e. $A_1=A_2$ and $A_3=A_4$, then the DOP for SEGSM beam is given by
\begin{equation}
P(\boldsymbol{\rho},\lambda)=|B_{xy}(\lambda)|.
\end{equation}
Thus we find that the degree of polarization for a SEGSM beam depends on the correlation coefficients between the orthogonal electric field components at different wavelengths and is independent from the Gaussian dependence of the spectral parameters and the source point $\boldsymbol{\rho}$. Clearly the DOP of the field indicates correlation between orthogonal field components, which is one (maximum correlation) for fully polarized field, zero (no correlation) for unpolarized field, and in between zero and one (partial correlation) for partially polarized light field. 

\section{Spectral degree of polarization for a 2-f system}
\begin{figure}[ht]
\centering
\includegraphics[width=0.9\linewidth]{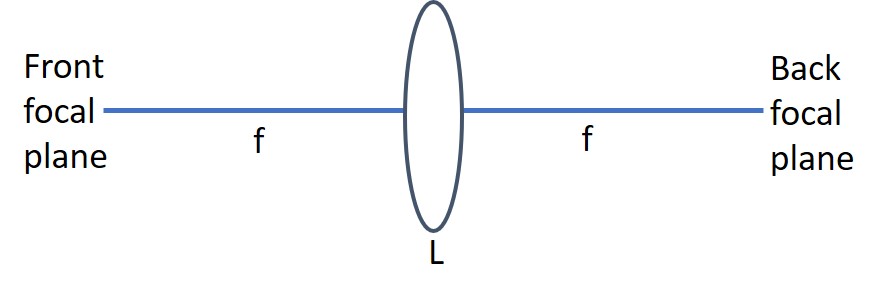}
\caption{Schematic diagram to study the degree of polarization using 2-f system. L is a lens having $f$ focal length.}
\end{figure}
Let us consider an SEGSM beam propagating through a 2-f lens system in which the back focal plane is a Fourier transform plane. If $W_{ij}(\boldsymbol{\rho}_1,\boldsymbol{\rho}_2,\lambda)$ is the element of CSD matrix at the front focal plane of 2-f lens system (shown in fig. 1), then using the tensor ABCD law \cite{wang02}, the element of CSD matrix at the back focal plane $W_{ij}(\boldsymbol{r}_1,\boldsymbol{r}_2,\lambda)$ of the lens can be obtained as
\begin{equation}
W_{ij}(\boldsymbol{r}_1,\boldsymbol{r}_2,\lambda)=\sqrt{S_i S_j}B_{ij}[det(\Bar A+\bar B M_1^{-1})]^{-1/2}exp[\frac{-ik}{2} \boldsymbol{r}^TM_p^{-1}\boldsymbol{r}], 
\end{equation}
where, $M_p^{-1}=(\Bar C+\bar DM_1^{-1})(\bar A+\bar BM_1^{-1} M_1^{-1})^{-1}$ and $ \boldsymbol{r}^T=(\boldsymbol{r}_1^T,\boldsymbol{r}_2^T)=(x_1,y_1,x_2,y_2)$ is 4*4 matrix, where notation $T$ stands for \textit{transpose} of the matrix. The values of matrices $\bar A$, $\bar B$, $\bar C$, $\bar D$  and $M_1^{-1}$ are given by \cite{wang02}
\begin{figure}[ht]
\centering
\includegraphics[width=0.9\linewidth]{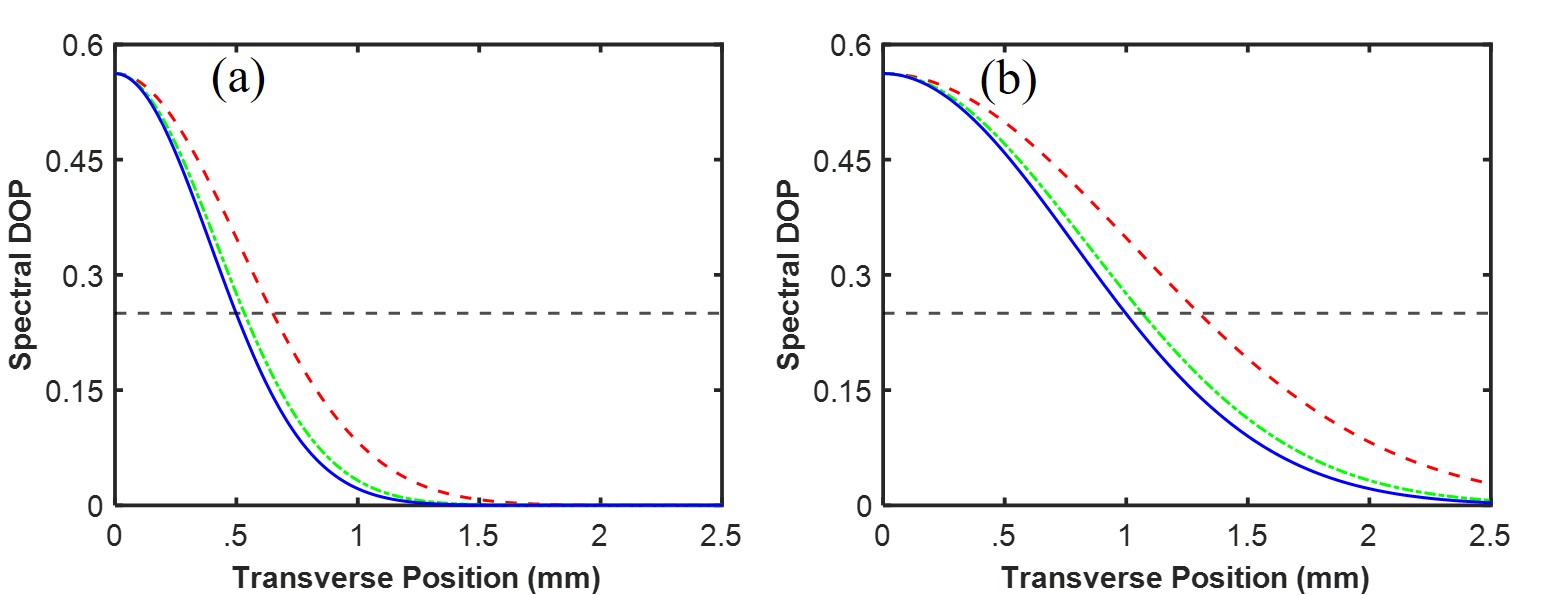}
\caption{Variation of spectral DOP of the output beam at the back focal plane of a 2-f lens system by varying the transverse distance from the optic axis of the beam for $\lambda$= central (peak) wavelength, $\lambda_S$=550nm, for (a) $f=1$ m, (b) $f=2$ m. In both (a) and (b), dash (red), dot dash (green) and solid (blue) lines are for $\lambda_n$( for n=$\delta$, $\delta_{xy}$)=450, 500, and 550 nm, respectively. The black horizontal dashed line shows the input beam DOP of 0.25.}
\end{figure}

\begin{subequations}
 \begin{equation}
\bar A=
\begin{bmatrix}
A & 0I\\0I & A^* 
\end{bmatrix},
\bar B=
\begin{bmatrix}
B & 0I\\0I & -B^* 
\end{bmatrix},
\bar C=
\begin{bmatrix}
C & 0I\\0I & -C^*
\end{bmatrix},
\bar D=
\begin{bmatrix}
D & 0I\\0I & D^*
\end{bmatrix},
\end{equation}
\begin{equation}
M_1^{-1}=
\begin{bmatrix}
\frac{-i}{k}[(\frac{1}{(2\sigma_\alpha^2)})+\frac{1}{\delta_{\alpha \beta}^2}]I & \frac{i}{k\delta_{\alpha \beta}^2}I \\\frac{i}{k\delta_{\alpha \beta}^2}I & \frac{-i}{k}[(\frac{1}{(2\sigma_\beta^2)})+\frac{1}{\delta_{\alpha \beta}^2}]I 
\end{bmatrix}, \text{for} (\alpha,\beta=x,y),
\end{equation}
\end{subequations}
and A=0I, B=fI, $C=\frac{-1}{f}I$, D=I, where I is a 2*2 unit matrix.\newline
\begin{figure}[ht]
\centering
\includegraphics[width=0.9\linewidth]{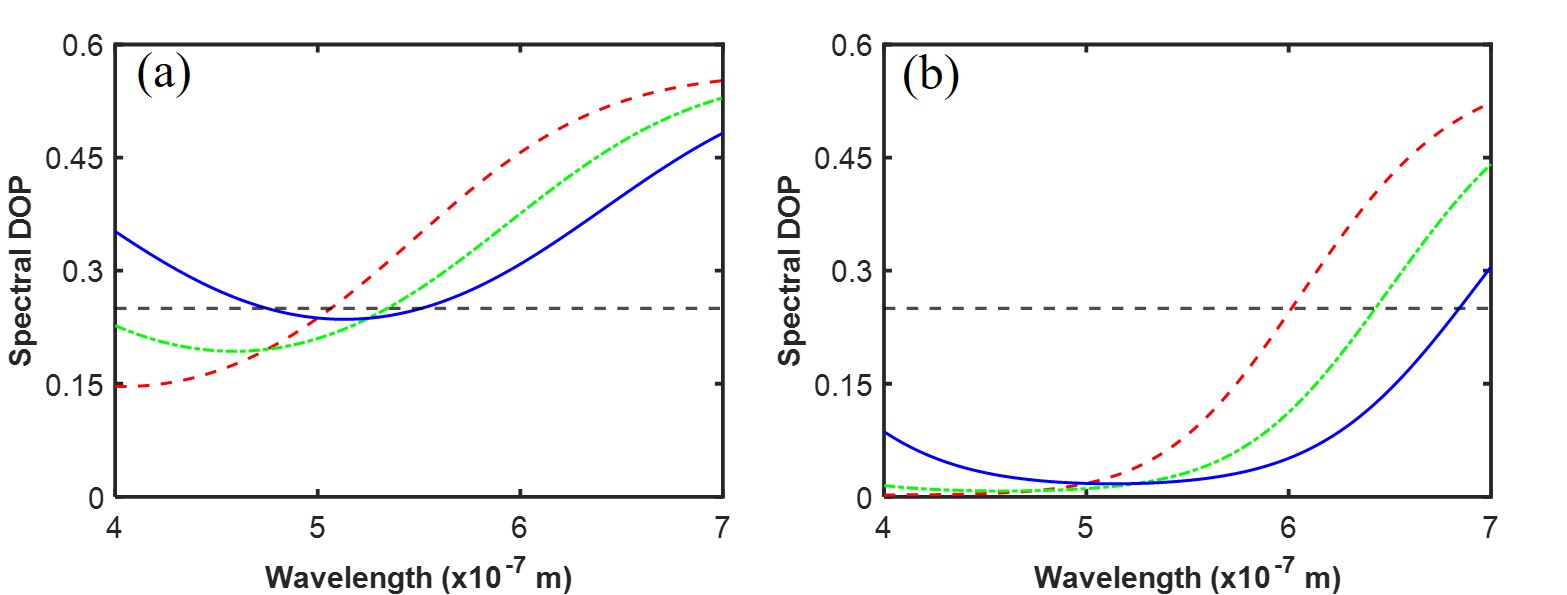}
\caption{Variation of spectral DOP of the output beam at the back focal plane of a 2-f lens system by varying the wavelength of the source for $f=1$ m, and transverse distance (a) $r=0.5 mm$, (b) $r=1 mm$. In both (a) and (b), dash (red), dot dash (green) and solid (blue) lines are for $\lambda_n$( for n=$\delta$, $\delta_{xy}$)=450, 500, and 550 nm, respectively. The black horizontal dashed line shows the input beam DOP of 0.25.}
\end{figure}
\begin{figure}[htbp]
\centering
\includegraphics[width=0.9\linewidth]{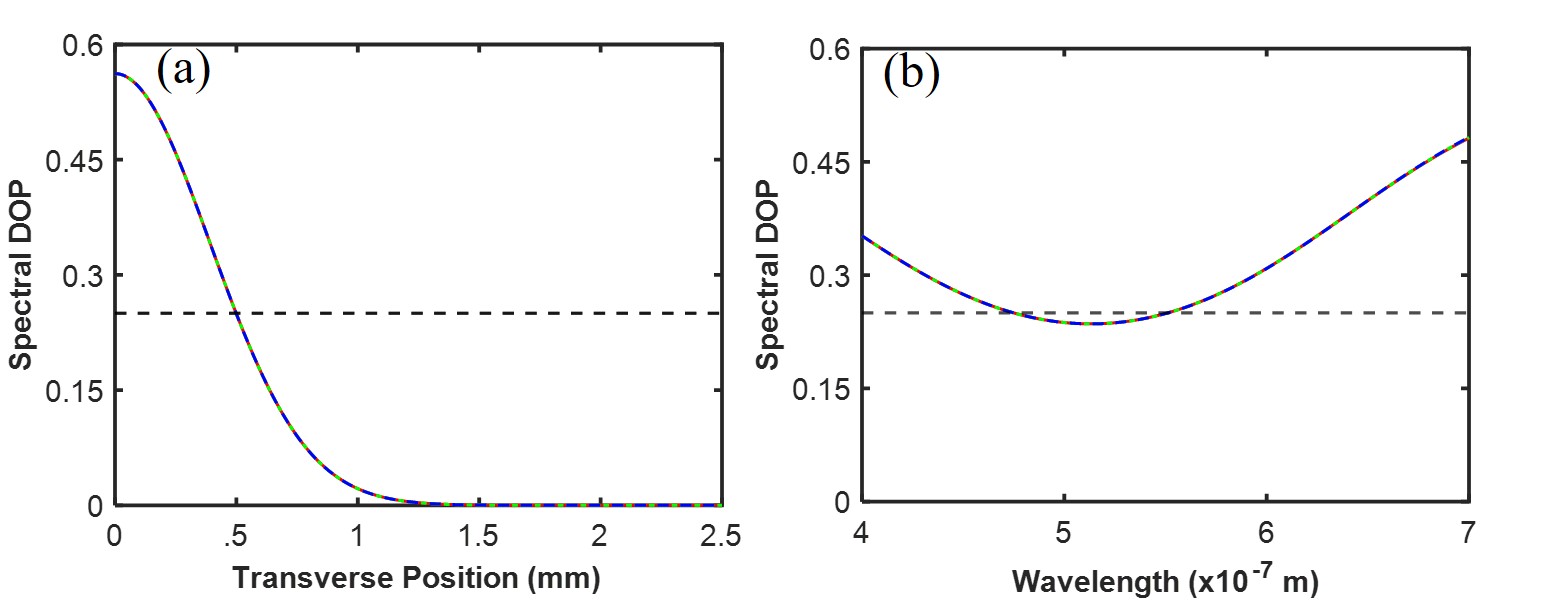}
\caption{Variation of spectral DOP of the output beam at the back focal plane of a 2-f lens system (a) by varying the transverse distance from the optic axis of the beam for $f=1$ m, (b) by varying the wavelength of the source for $f=1$ m, and transverse distance $r=0.5 mm$. In both (a) and (b), dash (red), dot dash (green) and solid (blue) lines are for $\lambda_\sigma$ =450, 500, and 550 nm, respectively (overlapping plots) and black horizontal dashed line shows the input beam DOP of 0.25. Spectral DOP remains unchanged for different values of $\lambda_\sigma$.}
\end{figure}
Using eqs. (12) (13) and (7), we find the spectral degree of polarization at the back focal plane of lens, which yields as
\begin{equation}
\begin{split}
P(\boldsymbol{r},\lambda)=\frac{\delta_{xy}^2(\lambda)}{\delta^2(\lambda)}\frac{\delta^2(\lambda)+4\sigma^2(\lambda)}{\delta_{xy}^2(\lambda)+4\sigma^2(\lambda)}|B_{xy}|exp[\frac{-8k^2r^2\sigma^4(\lambda)}{f^2}\\ \times {\frac{\delta_{xy}^2(\lambda)-\delta^2(\lambda)}{(\delta_{xy}^2(\lambda)+4\sigma^2(\lambda))(\delta^2(\lambda)+4\sigma^2(\lambda))}}].
\end{split}
\end{equation}
Since the SEGSM beam parameters $\delta_{xy}(\lambda)$, $\delta(\lambda)$, $\sigma(\lambda)$ follow the Gaussian distribution in spectral domain, using eqs. (5) and (14) we obtain DOP at the back focal plane of the lens, which depends on the spectral parameters of the Gaussian beam, transverse position of the beam at the back focal plane and wavelength of the source spectrum. The DOP is given by
\begin{equation}
P(\boldsymbol{r},\lambda)=\frac{C_{1}}{C_{2}}\frac{(C_{2}+C_{3})}{(C_{1}+C_{3})}|B_{xy}(\lambda)|C_{4},
\end{equation}
where the coefficients are given by 
$C_{1}=\delta_{xy(0)}^2exp\left[\frac{-(\lambda-\lambda_{\delta_{xy}})^2}{\Lambda_{\delta_{xy}}^2}\right]$,
$C_{2}=\delta_{0}^2exp\left[\frac{-(\lambda-\lambda_{\delta})^2}{\Lambda_{\delta}^2}\right]$,
$C_{3}=4\sigma_{0}^2exp\left[\frac{-(\lambda-\lambda_{\sigma})^2}{\Lambda_{\sigma}^2}\right]$, $C_4=exp[\frac{-C_5}{C_6}\frac{(C_1-C_2)}{(C_1+C_3)(C_2+C_3)}]$,  $C_5=32\pi^2r^2\sigma_0^4exp\left[\frac{-2(\lambda-\lambda_\sigma)^2}{\Lambda_\sigma^2}\right]$, and $C_6=\lambda^2f^2$.
\newline
From Eq. (15) we see that the DOP at the back focal plane of the lens depends on the spectral parameters of the SEGSM beam. The input beam parameters passing through 2-f lens configuration are fixed and taken as the following: peak wavelength of input spectra  $\lambda_S=550nm$, $\delta_{xy(0)}$=0.3mm, $\delta_{0}$=0.2mm, $\Lambda_{\delta_{xy}}$= $\Lambda_{\delta}$= $\Lambda_{\sigma}$=$\frac{\lambda_S}{4}$=137.5nm, $\sigma_{0}$=5mm, $\lambda_{\sigma}$=550nm, and $|B_{xy}|$=0.25. We choose the source parameters such that they satisfy the SEGSM beam realizability condition given in Eq. (9). We investigate the output beam for change in values of parameters (a) $\lambda_{\delta}$ and (b) $\lambda_{\delta_{xy}}$, provided all other beam parameters are fixed, and show that this can substantially affect the distribution of the spectral DOP of the propagating beam. In figure (2) we show the behaviour of spectral DOP of the output beam using Eq.(15) for peak wavelength of the source. We see that the spectral DOP decreases with the transverse position of the beam from the optic axis. With increase in the focal length of the beam, the decrease in DOP becomes less rapid. In figure (3) we show the change in spectral DOP using Eq.(15) for all wavelengths of the source. Firstly, the spectral DOP decreases with wavelength and after a particular wavelength given by the peak value of the beam spectrum, the spectral DOP increases. In figure (4), we show that the spectral DOP decreases with transverse position and its value is same for all value of $\lambda_\sigma$. In this case we have considered both $\lambda_\delta$, and $\lambda_{\delta_{xy}}$ same as 550nm. Since the back focal plane of lens receives contributions from every point in the front focal plane in 2-f arrangement and hence variation in DOP with position is expected. Thus in a nutshell, for a SEGSM beam passing through a 2-f lens system, the spectral DOP at the back focal plane depends on the transverse position of the beam, mean value of rms beam width and rms width of correlation function.

\section{Spectral degree of polarization in 4-f system}
\begin{figure}[ht]
\centering
\includegraphics[width=0.9\linewidth]{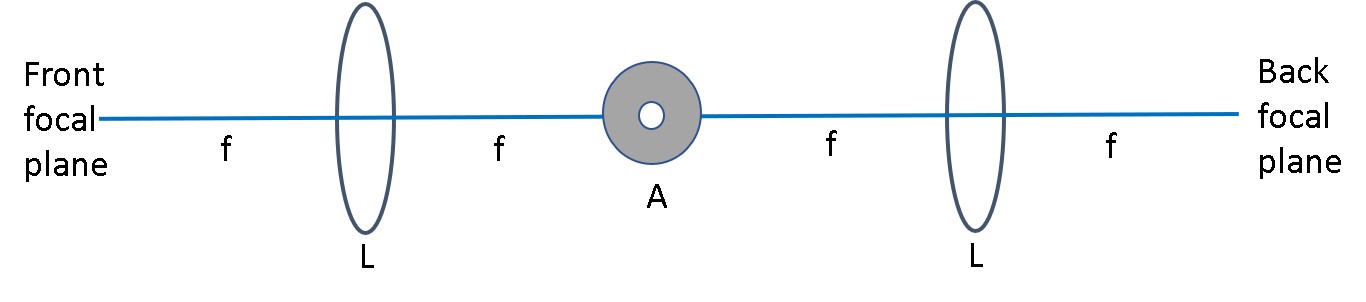}
\caption{Schematic diagram to study the degree of polarization using 4-f system. L is lens of f focal length and A is an aperture. }
\end{figure}
We next consider a 4-f lens system in which the back focal plane of the second lens gives the object plane. If a circular aperture having radius $a$ is placed between the two lenses of the 4-f system (figure 5), then at the back focal plane the DOP is controlled by the radius $a$ and focal length of the lenses \cite{zhao18}. For a SEGSM beam, in addition to these parameters the DOP can be controlled by the parameters $\lambda_{\delta}$ and $\lambda_{\delta_{xy}}$. The DOP of EGSM beam at the back focal plane of the 4-f system is given by \cite{zhao18},
\begin{figure}[ht]
\centering
\includegraphics[width=0.9\linewidth]{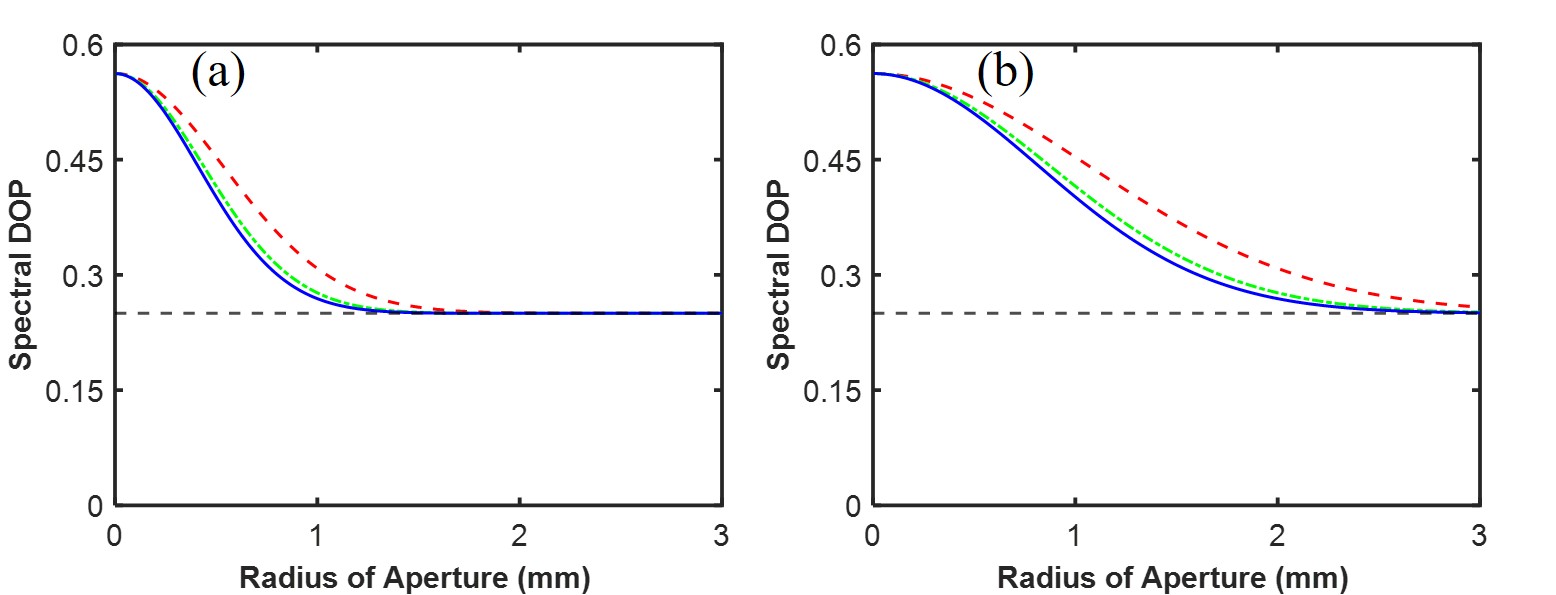}
\caption{(a) Variation of spectral DOP at the back focal plane of a 4-f lens system by varying the radius of aperture for (a) $f=1$ m, (b) $f=2$ m.  In both (a) and (b), dash (red), dot dash (green) and solid (blue) lines are for $\lambda_n$( for n=$\delta$, $\delta_{xy}$)=450, 500, and 550 nm respectively, and black horizontal dashed line shows the input beam DOP of 0.25.}
\end{figure}
\begin{figure}[ht]
\centering
\includegraphics[width=0.9\linewidth]{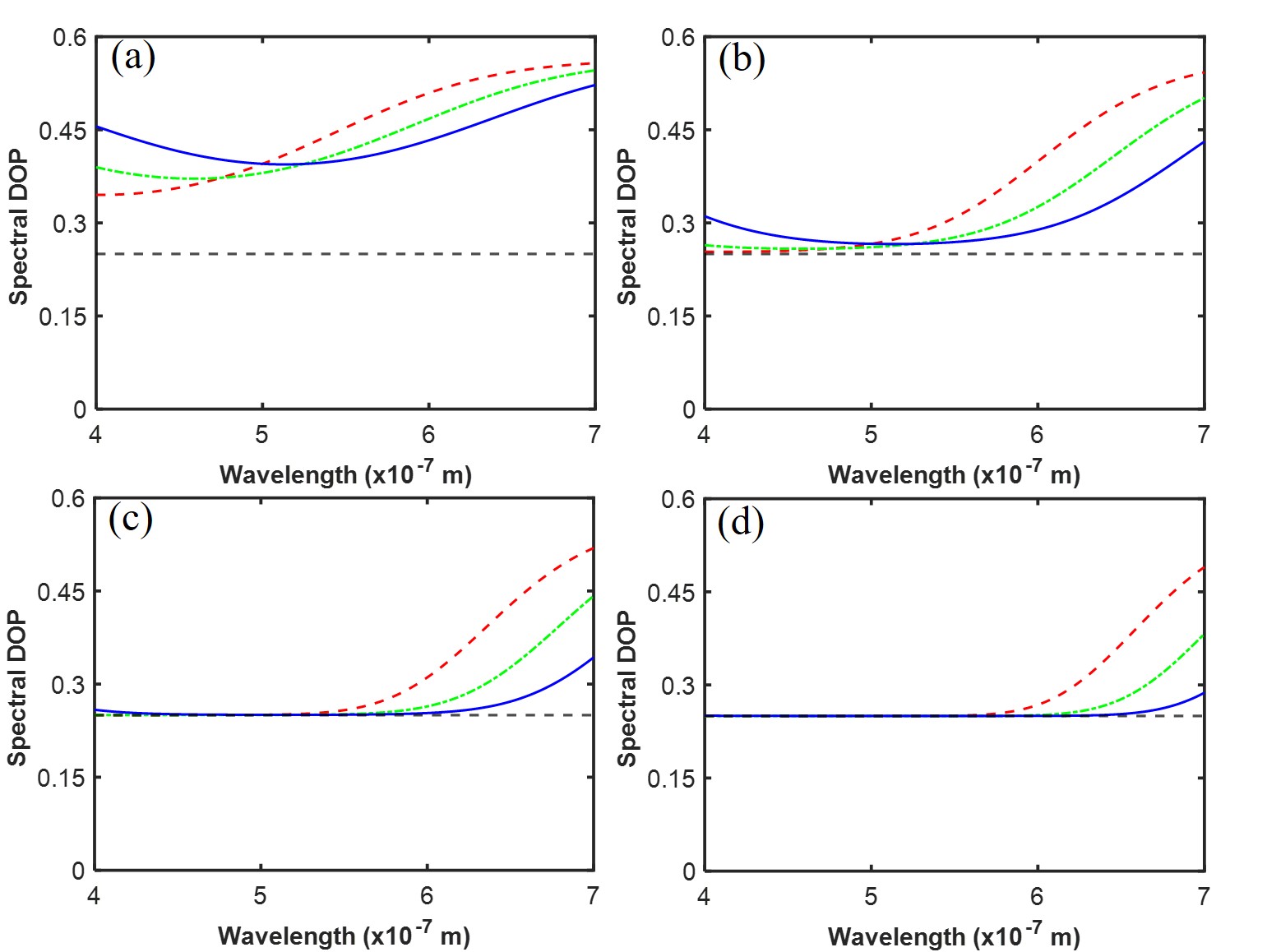}
\caption{Variation of spectral DOP at the back focal plane of a 4-f lens system by varying the size of the aperture  $a$ as (a) 0.5 mm, (b) 1mm, (c) 1.5mm, and (d) 2 mm, placed at the back focal plane of lens shown in figure 5.  In all four plots dash (red), dot dash (green) and solid (blue) lines are for $\lambda_n$( for n=$\delta$, $\delta_{xy}$)=450, 500, and 550 nm respectively, and black horizontal dashed line shows the input beam DOP of 0.25.} 
\end{figure}
\begin{figure}[ht]
\centering
\includegraphics[width=0.9\linewidth]{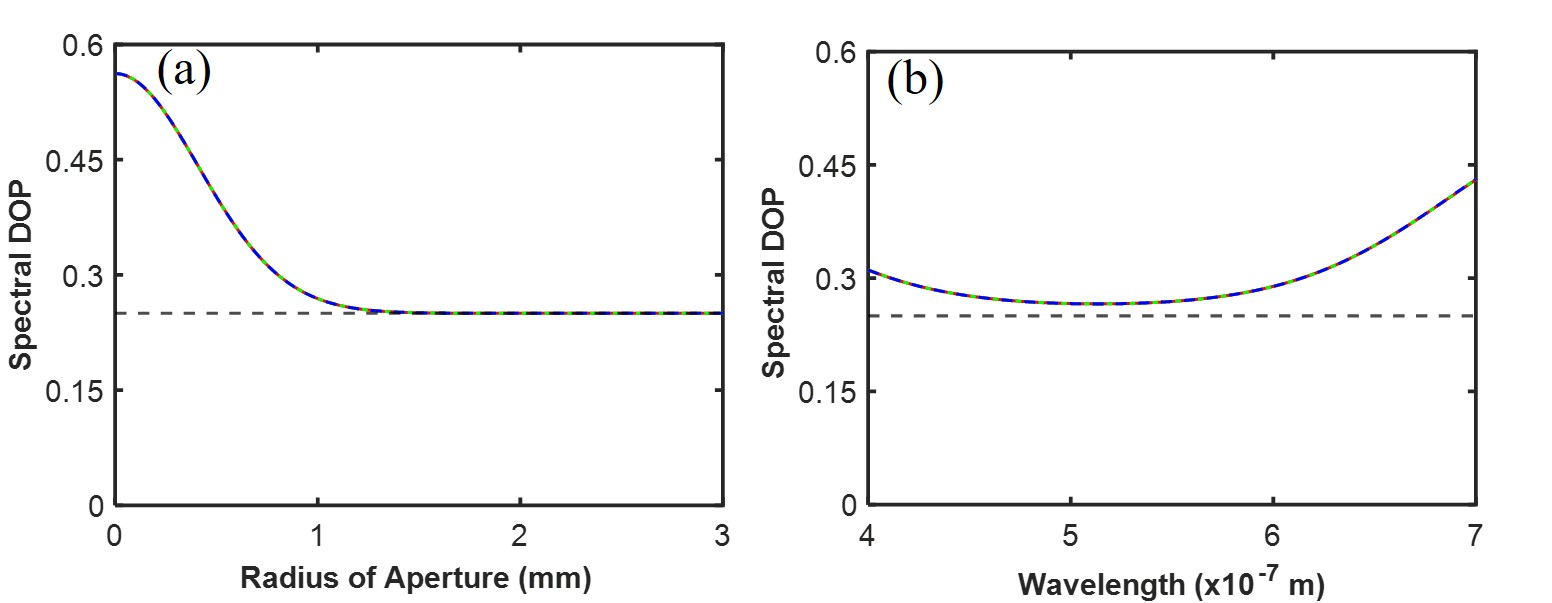}
\caption{Variation of spectral DOP of the output beam at the back focal plane of a 4-f lens system (a) by varying the radius of aperture for $f=1$ m, (b) by varying the wavelength of the source for radius of aperture $a=1$ mm and $f=1$ m, and transverse distance $r=0.5 mm$. In both (a) and (b), dash (red), dot dash (green) and solid (blue) lines are for $\lambda_\sigma$ =450, 500, and 550 nm, respectively (overlapping plots) and black horizontal dashed line shows the input beam DOP of 0.25. Spectral DOP remains unchanged for different values of $\lambda_\sigma$.}
\end{figure}
\begin{equation}
P(\lambda)=|B_{xy}|\left[\frac{1-exp(\frac{-k^2a^2}{2f^2}\frac{4\sigma^2\delta_{xy}^2}{\delta_{xy}^2+4\sigma^2})}{1-exp(\frac{-k^2a^2}{2f^2}\frac{4\sigma^2\delta^2}{\delta^2+4\sigma^2})}\right].
\end{equation} 
For SEGSM beam, the parameters $\sigma$, $\delta$ and $\delta_{xy}$ are given by Eq. (5). By substituting these values in Eq. (16), we get the expression for spectral DOP as 
\begin{equation}
P(\lambda)=|B_{xy}|\left[\frac{1-exp(-C_{7})}{1-exp(-C_{8})}\right],
\end{equation}
where we have the parameters 
$C_{7}=\frac{2\pi^2a^2(C_{3}C_{1})}{C_{6}(C_{1}+C_{3})}$,
and $C_{8}=\frac{2\pi^2a^2(C_{3}C_{2})}{C_{6}(C_{2}+C_{3})}$. Eq. (17) clearly shows that the output beam DOP is modulated by the spectral parameters of the beam. The spectral DOP at back focal plane of the second lens is found independent of the position of the beam. The reason for this independence may be the fact that field distributions are identical in 4-f system as source plane except of $180^0$ orientation change and hence no change in DOP with position is expected. It depends on the radius of aperture, the mean rms width of the beam and rms width of the correlation function.

The input beam parameters passing through the 4-f lens configuration are taken as $\delta_{xy(0)}$=0.3mm, $\delta_{0}$=0.2mm,$\Lambda_{\delta_{xy}}$=$\Lambda_{\delta}$=$\Lambda_{\sigma}$=137.5nm, $\sigma_{0}$=5mm, $\lambda_{\sigma}$=550nm, and $|B_{xy}|$=0.25, which satisfy the SEGSM beam realizability condition given in Eq. (9). In 4-f lens system, we use a circular aperture at the focal plane of first lens. Figure (6) denotes the spectral DOP of the output beam plotted using Eq. (17) for several peak wavelengths (spectrum) of the source. Spectral DOP decreases with the radius of aperture, and one can see that its minimum value is limited by the input beam DOP. This decreases in DOP is slower as the focal length of the lens system is increased. Figure (7) denotes the spectral DOP obtained using Eq. (17) for all wavelengths with increase in the aperture size. Since the aperture selects the spatial frequencies at the Fourier plane depending on its size, with bigger size more frequencies are allowed and correspondingly the DOP decreases and tends to the input DOP of the beam. Figure (8) shows the spectral DOP decreases with the radius of aperture and its value is same for any $\lambda_\sigma$ of the input beam. Thus in a nutshell, for a SEGSM beam passing through a 4-f lens system, the spectral DOP at the back focal plane depends on the spectral parameters of the beam.

\section{Conclusion}
In conclusion, we demonstrate theoretically the behaviour of the spectral degree of polarization of the SEGSM beam at the back focal plane of a 2-f and 4-f lens systems. At first we obtain a general expression of spectral DOP for the SEGSM beam having spectral density, rms beam width, and rms correlation function following the spectral Gaussian distribution. We find that the determined spectral DOP in the 2-f and 4-f lens configurations can be varied by changing the spectral parameters such as spectral rms width $\lambda_{\delta}$ and rms spectral correlation function $\lambda_{\delta_{xy}}$. It is observed that in the 2-f lens system, the spectral DOP of the central wavelength of the beam at the back focal plane of the lens changes with respect to the transverse position from the optic axis, whereas the spectral DOP for other wavelengths change significantly with respect to the spectral parameters of the beam. For a 4-f lens system, the spectral DOP at the back focal plane is independent of the transverse position of the beam, whereas it depends on the mean value of rms beam width, rms width of correlation function and size of aperture placed at the Fourier plane of the lens system. This study clearly demonstrates that the polarization properties of random statistically stationary electromagnetic beam like fields on focusing depend on their spectral content, which varies significantly at the source plane (4-f system) and at the image plane (2-f system).

\section{Acknowledgement} 
We thankfully acknowledge the funding received from Science and Engineering Research Board (SERB), India through grant YSS/2015/000743, and from Council for Scientific and Industrial Research (CSIR), India through grant 03(1401)/17/EMR-II. Author RJ is thankful to CSIR, India for a Senior Research Fellowship.

\end{document}